\documentclass[twocolumn,aps,prc,superscriptedaddress,reprint,floatfix]{revtex4-1}
\usepackage{graphicx}
\usepackage[latin2]{inputenc}
\usepackage[T1]{fontenc}
\usepackage{wrapfig}
\usepackage{titlesec}
\usepackage{gensymb}
\usepackage{amsmath}
\usepackage[caption=false]{subfig}
\usepackage{tabularx}

\makeatletter
\gdef\@ptsize{0}% 10pt documents
\let\@currsize\normalsize
\makeatother
\usepackage{setspace}

\begin{document}
\titlespacing\section{0pt}{20pt plus 4pt minus 2pt}{3pt plus 20pt minus 2pt}
\titleformat{\section}
  {\normalfont\fontsize{10}{10}\bfseries\centering}{\thesection}{0.5em}{}
\renewcommand\thesection{\Roman{section}.}

%*** ADD YOUR TITLE HERE ***%
\title{\bf Systematic investigation of channel coupling effects on elastic, inelastic and neutron transfer channels in $^6$Li+$^{159}$Tb}
\author{Saikat Bhattacharjee$^{1,3}$}
\author{Piyasi Biswas$^{1,3}$}
\altaffiliation{Present address: Shahid Matangini Hazra Government College for
Women, Tamluk, Chakshrikrishnapur, Kulberia, Purba Medinipur,
West Bengal - 721649}
\author{Ashish Gupta$^{1,3}$}
\author{M. K. Pradhan$^{1}$}
\altaffiliation{Present address: Department of Physics, Belda College, Belda,
Paschim Medinipur, West Bengal - 721424}
\author{N. Deshmukh$^{1}$}
\altaffiliation{Present address: School of Sciences, Auro University, Surat, 
Gujarat - 394510, India}
\author{P. Basu$^{1}$}
\altaffiliation{Retired}
\author{V. V. Parkar$^{2,3}$}
\author{S. Santra$^{2}$}
\author{K. Ramachandran$^{2}$}
\author{A. Chatterjee$^{2}$}
\author{Subinit Roy$^{1}$}
\altaffiliation{Retired}
\author{A. Mukherjee$^{1,3}$} %PRESENTATION AUTHOR
\altaffiliation {anjali.mukherjee@saha.ac.in}
\affiliation{$^1$Saha Institute of Nuclear Physics, 1/AF, Bidhan Nagar, Kolkata-700064, India} 
\affiliation{$^2$Nuclear Physics Division, Bhabha Atomic Research Centre, Mumbai-400085, India}
\affiliation{$^3$Homi Bhaba National Institute, Mumbai-400094, India}
\begin{abstract}
Elastic scattering angular distribution for weakly bound nucleus $^{6}$Li on the deformed rare earth $^{159}$Tb target nucleus has been measured at energies around the Coulomb barrier. The elastic scattering cross sections for this reaction consist of inelastic contributions from low lying excited states of $^{159}$Tb. The pure elastic cross-sections have been extracted from the admixture of elastic and inelastic data. The optical model potential parameters for the system have been obtained from the extracted pure elastic scattering cross sections. Coupled channel calculations have been performed with this set of potential parameters, to compare the theoretical and experimental inelastic scattering cross sections. The work has been extended to obtain the spectroscopic factor for $^{158}$Tb+n configuration from the experimental 1n-pickup data. 
\end{abstract}

%*** DO NOT CHANGE NEXT COMMAND ***%
\maketitle

\section{Introduction}
Reaction process between two heavy ions has been studied extensively over the last few decades at energies around the Coulomb barrier to get a deeper understanding of the  nuclear potential. In recent years, the study of the model nuclear potential has been focused on stable weakly bound nuclei at energies around Coulomb barrier \cite{Pakou03,Pakou04,Pakou08,Keeley94,Maciel99,Lubian07,Hussein06,Gomes05,Moraes00,
Woolliscroft04,Gomes04,Gomes05a,Gomes06,Figueira06,Figueira07,Figueira10,Camacho08,
Kumawat08,Signorini00,Zardo09,Biswas08,Aguilera09,Garcia07,Yu10,Deshmukh11,Patel15,
Yang16,Mazzocco19,Kumawat20}. Because of their prevalent cluster structure, nuclear reactions involving the weakly bound nuclei have increased probability of breakup and transfer \cite{Canto06}. The systematics of such increased transfer and breakup probability influences the elastic scattering cross-sections \cite{Gomes16}. Consequently, the mean field potential extracted from the elastic scattering measurement is also affected due to the increased probability of breakup and transfer, reflected in the energy dependence of optical model potential. The near threshold behaviour of the potential, known as Threshold Anomaly (TA) \cite{Satchler91} exhibits a different behaviour for weakly bound projectiles , unlike the strongly bound systems \cite{Hussein06,Gomes05}.  

Highly deformed rare earth nuclei \cite{Aponick70} like $^{159}$Tb, have large density of excited states adjacent to the ground state. Elastic scattering measurements involving these nuclei generally yields the quasi-elastic data, admixture of elastic and inelastic scattering to low-lying states. Subsequently, a statistical model can be implemented to extract the elastic cross section from the quasi-elastic data \cite{Birkelund76}.

Measurements of fusion \cite{Pradhan11}, $\alpha$-yield \cite{Pradhan13}, and quasi-elastic barrier distribution \cite{Biswas21} have been reported for $^{6}$Li+$^{159}$Tb system. However, the elastic scattering measurement for $^{6}$Li+$^{159}$Tb does not exist in literature, and study of $^7$Li+$^{159}$Tb \cite{Patel15} exhibits unusual energy dependence of optical model potential parameters.  In that scenario, elastic scattering measurement for weakly bound $^6$Li projectile on  permanently deformed rare earth nucleus $^{159}$Tb  as target has been presented in this work. The work further extends to probe the one neutron (1n) pickup reaction of $^{6}$Li from $^{159}$Tb. 

Section II of this paper consists of the experimental details, analytical procedures to extract the elastic part from quasi-elastic data and the search for optical model potential parameters. Section III recounts the theoretical calculations that have been performed to reproduce the quasielastic and transfer data at different energies. Summary of the present work and concluding remarks are included in Section IV.

\section{Experimental Details and Analysis}
\subsection{Experimental details}
The experiment has been performed at  14UD BARC-TIFR Pelletron Accelerator at TIFR, Mumbai. $^{6}$Li beam with lab energies 25, 27, 30, 35 MeV has been used to bombard a self supporting rolled target foil of $^{159}$Tb with thickness of 700 $\mu g/cm^2$. Four Silicon (Si) surface barrier $\Delta$E-E telescope detectors were employed in an angular range 20$\degree \leq \theta_{lab} \leq$160$\degree$ to detect the scattered particles at different angles. Thickness of the four $\Delta$E-E detectors are: 25 $\mu m$- 500 $\mu m$; 25 $\mu m$- 2000 $\mu m$; 25 $\mu m$- 3000 $\mu m$; and 25 $\mu m$- 2000 $\mu m$, respectively. Two monitor detectors were placed at $\pm$10$\degree$ with respect to beam direction for normalizing purposes. A typical $\Delta$E-E$_{tot}$ spectrum (E$_{tot}$ corresponds to the sum of energies deposited in $\Delta$E and E detectors)  is shown in Figure 1. The  $^{6}$Li band in Figure 1 is considered as the elastic band. At the top of the $^{6}$Li, formation of a $^{7}$Li band represents the 1n pickup channel. The Linux based data acquisition system LAMPS \cite{Chatterjee08} has been employed to register the data. 
\subsection{Extraction of elastic scattering data}
The $^{159}$Tb nucleus is a highly deformed with an unpaired proton coupling with the 0$^+$, 2$^+$,... pure rotational states. As a consequence, $^{159}$Tb has low lying excited states at 58 keV, 137 keV and so on. The Si surface barrier detectors used at the experiment have an estimated energy resolution of $\approx$ 150 keV. Hence, contribution of inelastic scattering by the low lying excited states of the target can not be separated by experimental means. Thus, the largest blob in Li band (see Figure 1) will consist of contribution from inelastic scattering. It is therefore requisite to separate the pure elastic data from the admixture of elastic and inelastic data. In Figure 2, the measured quasi-elastic angular distribution cross sections at energies E$_{lab}$ = 25, 27, 30 and 35 MeV have been presented. The factor $\mathrm{\sigma_{qel}/\sigma_{Ruth}}$ in Figure 2 represents the quasi-elastic differential scattering cross-sections normalized to Rutherford cross-section.
\begin{figure}[t]
	\centering
    \includegraphics[clip, trim=1.2cm 3.5cm 1.0cm 1.5cm,width=0.46\textwidth]{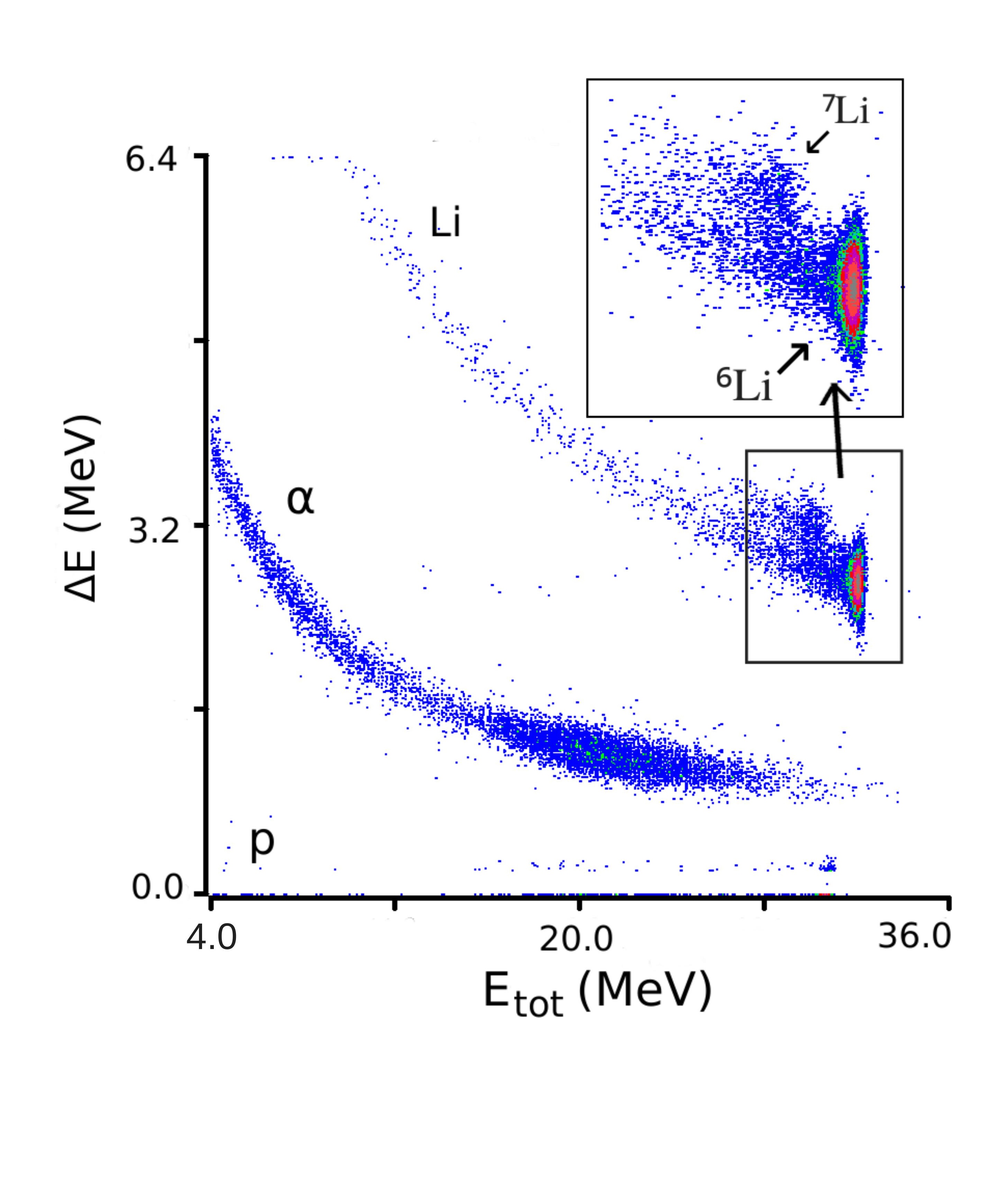}
  	\caption{Typical $\Delta$E-E$_{tot}$ spectra for scattering of $^{6}$Li+$^{159}$Tb system at E$_{lab}$ = 35 MeV and  $\theta_{lab}=59.5\degree$. The uppermost band, with an enlarged view in the inset; consists of scattered $^{6}$Li projectile, either elastically or inelastically; and $^{7}$Li emerging from 1n pickup of $^{6}$Li.}
\end{figure} 

A one dimensional projection of Li band is shown in Figure 3. The peak is fitted by a statistical double gaussian fitting method to disentangle the elastic data (see Figure 3). The fitting was done with the following constraints:

(1) The peak of the second gaussian will be at an energy which is 58 keV less than the peak of elastic scattering. The first excited state of $^{159}$Tb is 58 keV above the ground state. Imposing this condition takes care of the peak position of 1st inelastic state.

(2) FWHM of the two peaks of the double gaussian will be same. This condition comes from the fact that energy resolution of the detector, which is roughly equal to the FWHM of the peaks does not change significantly over a  difference of 58 keV in energy.

(3) Ratios of elastic to inelastic cross-section at different angles have been estimated from the initial coupled channel calculations employing different global potentials \cite{Cook82, Xu18}. The experimental single particle transition strengths for $^{159}$Tb have been incorporated in the calculations to fix the ratio. During the fitting procedure, the ratios have been maintained to obtain a more reliable result.  

It should be noted that contribution from the second inelastic state of $^{159}$Tb; situated at 137 keV above the ground state, is also present within the data. But including a third gaussian in the fitting algorithm was not feasible as the gaussian peaks could not be distinguished. Furthermore, including another gaussian would increase the number of free parameters, rendering the fitting less accurate; as different combination of the parameters may yield same result. Thus, the extracted elastic and inelastic parts probably still consist of contribution from 137 keV inelastic state. The fitting has been performed with the aid of $\chi^2$ minimization method. The technique has been successfully carried out at above barrier energies at scattering angles where the ratio of inelastic to elastic scattering cross section is greater than $3-4\%$. At energies below Coulomb barrier, the ratio of inelastic and elastic scattering cross section becomes smaller ($<3-4\%$) even at higher angles which can not be resolved statistically. The attribute becomes more prominent with decreasing energies. In these circumstances, the total cross sections estimated from the raw data have been considered to come from elastic scattering only. Errors in the fitted parameters as well as the statistical error of the data have been taken into account while calculating the overall error for either elastic or inelastic component. As the errors of the data points include error from the double gaussian fit along with the statistical error of counts, the error bars associated with the data points are higher than that usual.
\begin{figure}[t]
	\centering
    \includegraphics[clip, trim=0.0cm 0.0cm 0.0cm 0.0cm,width=0.49\textwidth]{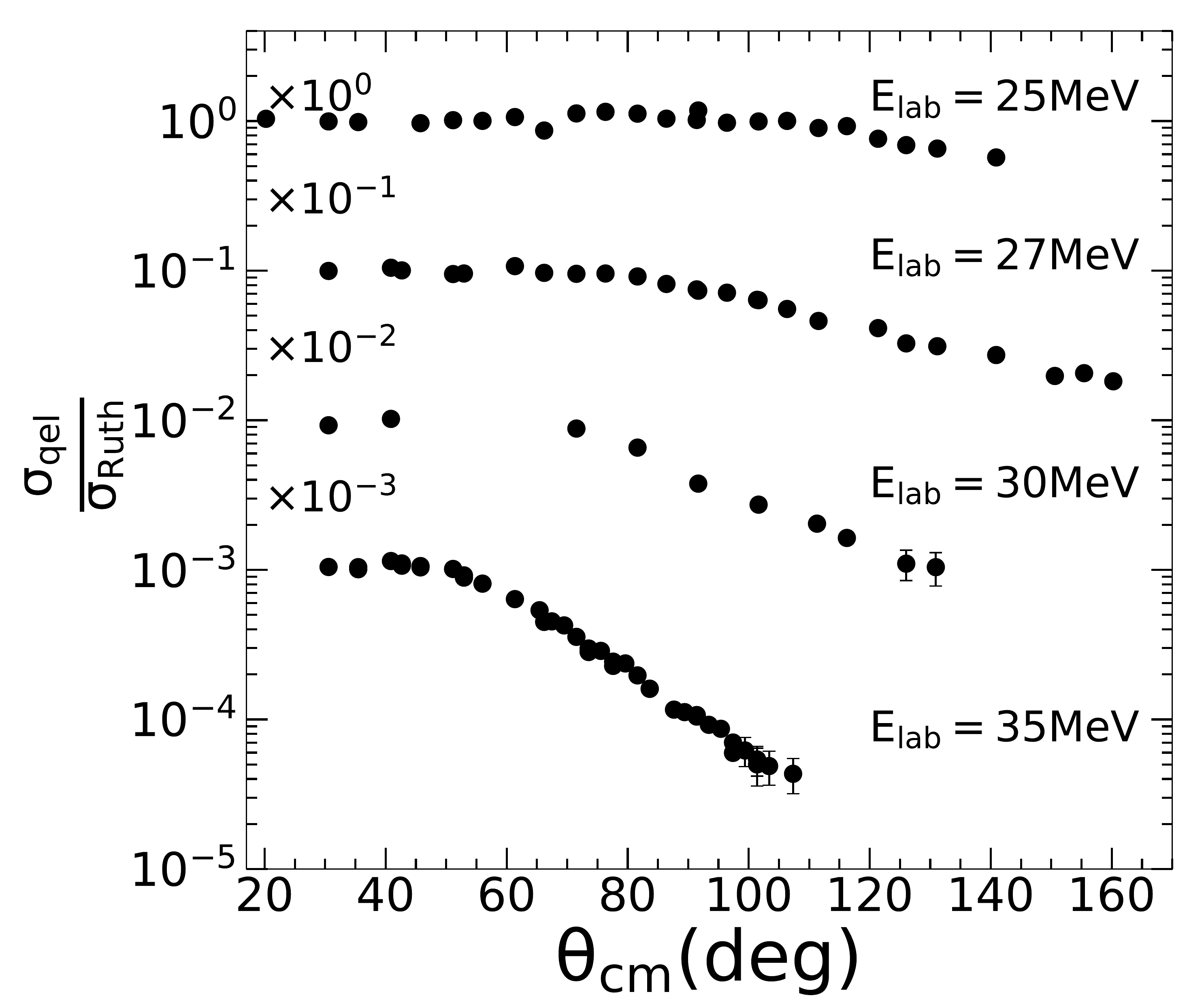}
  	\caption{Experimental angular distribution of quasielastic scattering cross-sections  at different incident energies for $^6$Li+$^{159}$Tb system. }
\end{figure} 

\begin{figure}[t]
	\centering
    \includegraphics[clip, trim=0.0cm 0.0cm 0.0cm -0.4cm,width=0.48\textwidth]{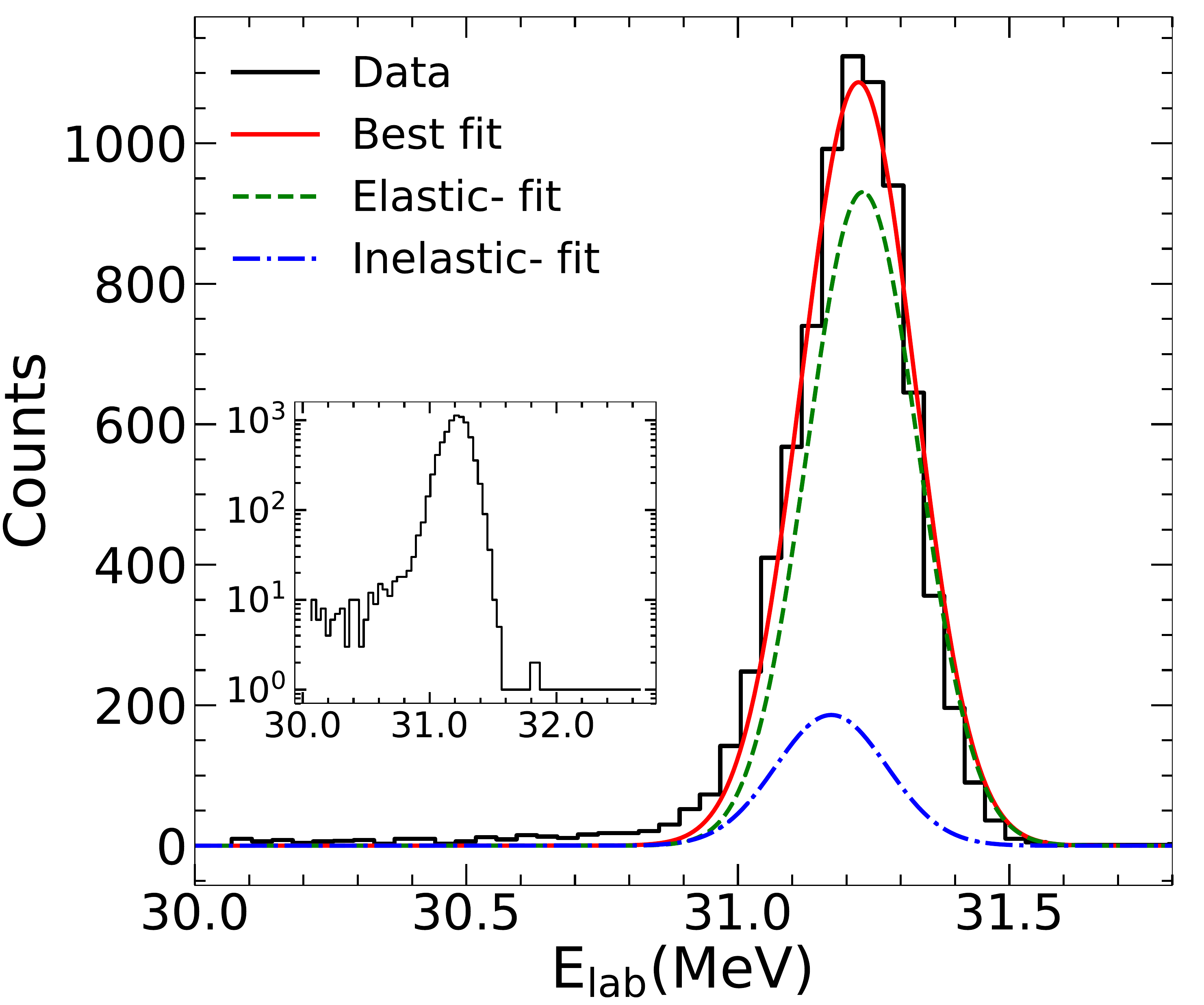}
  	\caption{Result of double gaussian fit performed on the elastic+inelastic data (at ${\theta _{lab}=75.5^{\degree}}, E_{lab}$ = 35 MeV ). The elastic data extracted (dashed lines) from the fit has been used to obtain the  model potential parameters of the system. (inset) Deviation of the data from the gaussian shape can be observed, which indicates an admixture of elastic and inelastic cross-section. }
\end{figure}  

The ratio of elastic to Rutherford differential cross-section is then expressed as:
\begin{align*}
\frac{d\sigma_{el}}{d\sigma_{Ruth}}(E,\theta_{tel})=\frac{Y_{el}(E,\theta_{tel})}{Y_m(E,\theta_m)}\times\\
\frac{(d\sigma_{Ruth}/d\Omega)(E,\theta_m)}{(d\sigma_{Ruth}/d\Omega)(E,\theta_{tel})}
(\frac{\Delta\Omega_m}{\Delta\Omega_{tel}})\tag{1}
\end{align*}
where, Y$_{el}$ is the yield of elastic scattering only, obtained from the double gaussian fit and Y$_m$ is the average yield of monitor detectors. The factor ${(d\sigma_{Ruth}/d\Omega)(E,\theta_m(\theta_{tel}))}$ corresponds to the differential Rutherford scattering cross section for a particular beam energy E, at a monitor angle $\theta_m$(or telescope detector $\theta_{tel}$). $\frac{\Delta\Omega_m}{\Delta\Omega_{tel}}$ is the solid angle ratio of monitor and telescope detector. The ratio has been obtained independently for four $\Delta$E-E telescope detectors from the lowest forward angles (30$\degree$-40$\degree$) of the lowest beam energy of 23 MeV, where the scattering is purely Rutherford. The mentioned energy is well below the Coulomb barrier of the system and therefore the elastic scattering in this energy is entirely Rutherford scattering, especially at lower angles.
\subsection{Search for the model potential parameters}
 The calculated ratios of elastic to Rutherford differential scattering cross-sections have been used as the input data in the search code SFRESCO \cite{Thompson88}. The initial set of potential parameters are derived from the global potential parameters of $^{6}$Li, determined by J.Cook \cite{Cook82}. The resultant potential parameters are: $V_0$= 109.5 MeV; $r_V$= 1.326 fm; $a_V$= 0.811 fm; $W_0$= 24.96 MeV; $r_W$= 1.534 fm and $a_W$= 0.884 fm.
  \begin{table}[t]
\caption{Optical model potential parameters correspondig to best fit at different energies for $^6$Li+$^{159}$Tb system. The potential is of Wood-Saxon form with total potential given by $V+iW$. $\chi^2/n$ is the reduced $\chi^2$ where $n$ is the number of data points at each energy.} % title of Table
\centering % used for centering table
\begin{tabular*}{0.49\textwidth}{c @{\extracolsep{\fill}} ccccccccc} % centered columns (7 columns)
\hline\hline \\[-2.0ex]%inserts double horizontal lines
E$_{lab}$ & $a_V$ & $a_W$ & $r_V$ & $r_W$ & $V_0$ & $W_0$ & $\chi^2/n$ & $\sigma_R$\\
(MeV) &  (fm) & (fm) & (fm) & (fm) & (MeV) & (MeV) &  & (mb) \\[1.5ex] % inserts table
%heading
\hline % inserts single horizontal line
%23 & 0.8 & 0.8 & 1.09 & 1.1 & 0   \\ % inserting body of the table
25 & 0.8 & 0.8 & 1.09 & 1.1 & 78$\pm$14 & 110$\pm$11 & 5.507 & 632.65   \\
27 & 0.8 & 0.8 & 1.09 & 1.1 & 77$\pm$6 & 122$\pm$7 & 1.43 & 893.13   \\
30 & 0.8 & 0.8 & 1.09 & 1.1 & 75$\pm$7 & 95$\pm$12 & 1.023 & 1164.63   \\
35 & 0.8 & 0.8 & 1.09 & 1.1 & 66$\pm$3 & 82$\pm$6 & 2.98 & 1537.84  \\ [1.0ex] % [1ex] adds vertical space
\hline %inserts single line
\end{tabular*}
%\label{table:t1} % is used to refer this table in the text
\end{table}

\begin{figure}[t]
\centering
\includegraphics[clip, trim=0.4cm 0.2cm 0.0cm 0.0cm,width=0.49\textwidth]{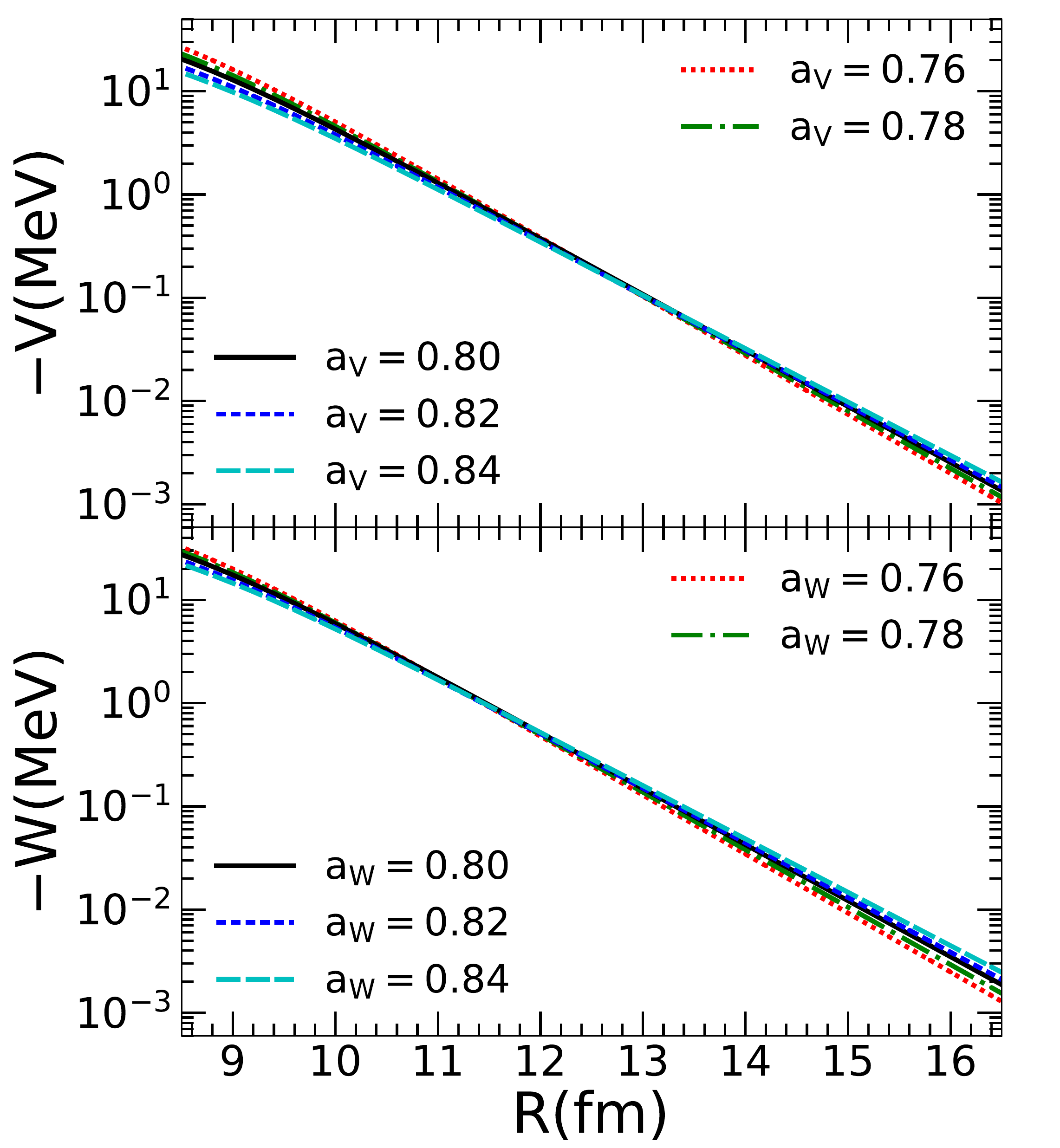}
\caption{Sets of different potential parameters that can produce good fit results (shown for $E_{lab}=35$ MeV. Intersection of the potentials generated by the obtained set of potential parameters are the estimated $R_{Sr}$(a) and $R_{Si}$(b). See text for further explanation.}
\end{figure} 

The highest energy data at $E_{lab}$ = 35 MeV was chosen to optimize the fitting procedure. In order to avoid the fitting with a large number of parameters at a time, the radius and diffusivity of real (r$_V$,a$_V$) and imginary (r$_W$,a$_W$) potentials were kept fixed while the strengths of the volume potentials were varied to obtain the best fit.  Afterwards, grid search on the diffusivity parameters were performed within a range from 0.74-0.84 fm to observe a minimum in reduced $\chi^2$ value. Several sets of potential parameters provided good fit, among which, the the optimum reduced  $\chi^2$=  2.98 was considered as the best fit result. The values of sensitivity radii $R_{Sr}$= 12.69 fm and $R_{Si}$= 11.45 fm for real and imaginary parts, respectively; have been calculated from the intersection of different potential parameter sets and shown in Figure 4.  Average value of $R_{Sr}$ and $R_{Si}$, i.e. 12.07 fm has been considered as the mean sensitivity radius.

The potential parameters at other energies, both below and above Coulomb barrier, have been obtained by following similar procedure. Parameters corresponding to the best fit at each energy has been provided in Table 1. The values of $a_i$ and $r_i$ ($i$= real, imaginary) are kept fixed for all energies.   The optical model potential has exhibited an energy dependence that is evident from the values in Table 1.  But, the nature of the energy dependence could not be investigated explicitly due to the unavailability of sufficient angular distribution data at more energies. The angular distribution of elastic scattering cross-section  at different energies along with the best fit obtained from SFRESCO has been shown in Figure 5. 
\begin{figure}[t]
	\centering
    \includegraphics[clip, trim=0.0cm 0.1cm 0.0cm 0.0cm,width=0.47\textwidth]{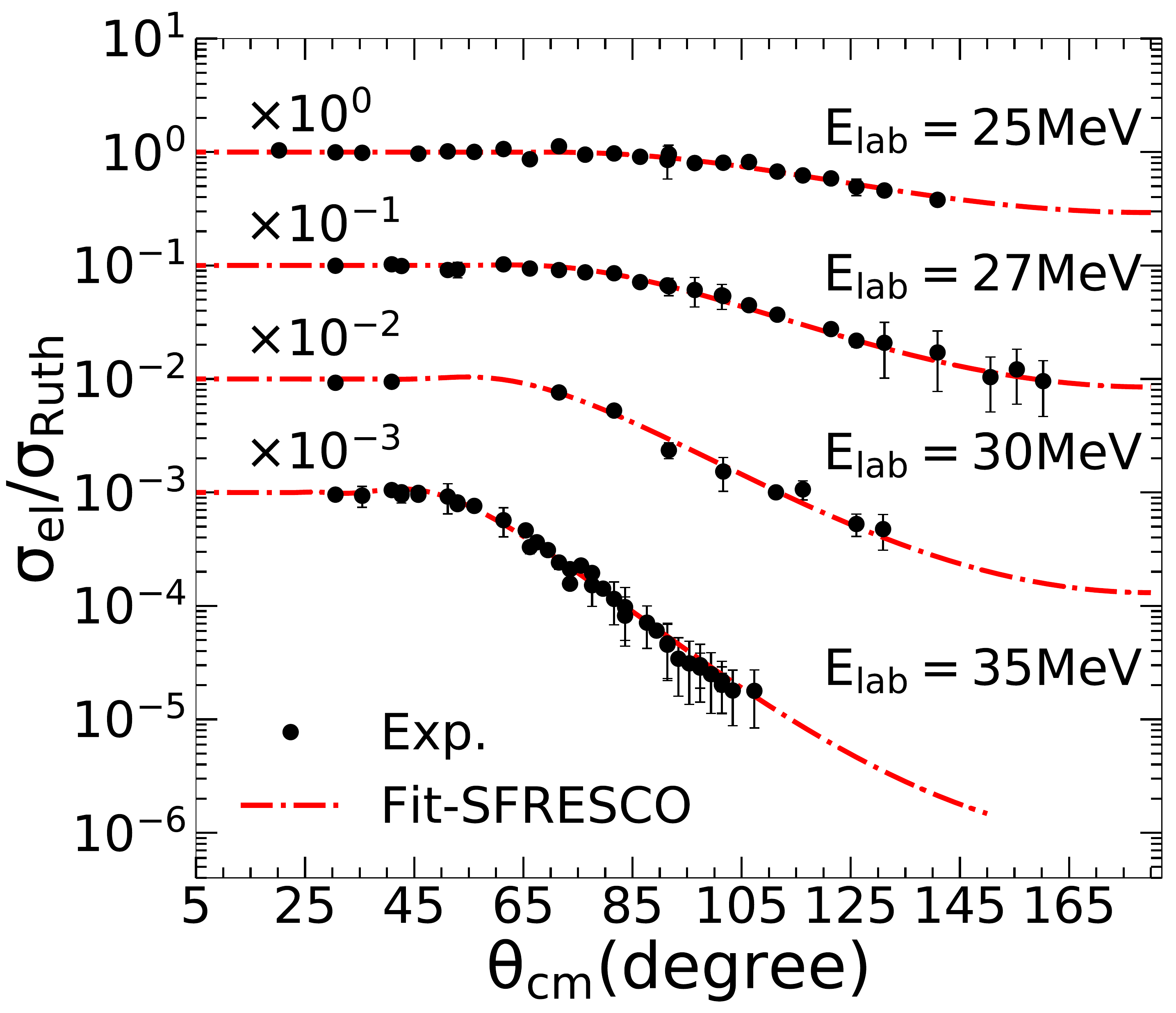}
  	\caption{Extracted angular distribution of differential elastic scattering cross-section ($\mathrm{\sigma_ {el}/\sigma_{Ruth}}$) at different energies below and above Coulomb barrier. The best fit, corresponding to minimum reduced $\chi^2$, has been obtained from fit with SFRESCO.  }
\end{figure}

\section{Theoretical Calculations}
\subsection{Coupled Channel Calculation}
 So far, the statistical estimation of purely elastic contribution generated by average interaction potential of $^{6}$Li+$^{159}$Tb has been considered for phenomenological calculations. However, it is important to probe the reliability of the model potential parameters thus obtained. For this purpose, the total contributions coming from elastic as well as inelastic states of $^{159}$Tb have been compared with coupled channel (CC) calculation, using the phenomenologically obtained potential parameters. It is to be mentioned that the quasielastic scattering cross-section does not include the contribution from $^7$Li (see Figure 1). Moreover, reproduction of the quasi-elastic data is necessary to justify the process of extracting the elastic scattering cross-section from quasi-elastic scattering data. 

The coupled channel (CC) calculation has been perofrmed by coupling the ground state (g.s.) and the first two excited states (e.s) of the target $^{159}$Tb with ground state of projectile $^{6}$Li \cite{Tilley02,Reich12} (see Table II), thereby calculating the elastic as well as the inelastic contribution due to first two e.s. of $^{159}$Tb. The excitation of projectile has not been included within the calculation as $^6$Li has no other bound state. The results of CC calculation have been compared with experimental quasielastic scattering angular distribution. The CC calculation has been performed using the  code FRESCO \cite{Thompson88}.
\begin{table}[t]
    \caption{Spin-parity ($J^\pi$) and energy of states for $^6$Li+$^{159}$Tb system which has been used for calculation.}
    \begin{tabularx}{1.0\columnwidth}{XXX|XXX}
        \hline\hline\\ [-2.0ex]
     	& \textbf{$^6$Li} &	&	& \textbf{$^{159}$Tb} &  \\[0.6ex] 

        $J^\pi$ &	& E (keV)	& $J^\pi$ &	& E (keV) \\[0.6ex]
        \hline
        1$^+$ & & $\ \ $g.s. & $\frac{3}{2}^+$ & & $\ \ $g.s. \\[0.7ex]
        & & & $\frac{5}{2}^+$ & & $\ \ $57.99 \\[0.7ex]
        & & & $\frac{7}{2}^+$ & & $\ \ $137.5  \\[0.7ex]
        & & & $\frac{9}{2}^+$ & & $\ \ $241.5  \\[1.0ex]
        \hline
    \end{tabularx}
\end{table}

\begin{table}[h]
\caption{Values of inelastic transition elements for $^{159}$Tb that has been used for calculation. $I_f$ and $I_i$ are the spins of final and initial states, respectively. E$_\gamma$ is the $\gamma$-ray energy for transition between two states.  } % title of Table
\centering % used for centering table
\begin{tabular*}{0.48\textwidth}{c @{\extracolsep{\fill}} cccc} % centered columns (7 columns)
\hline\hline \\[-2.0ex]%inserts double horizontal lines
$I_f \leftrightarrow I_i$ & E$_\gamma$(keV) & B(E2)(W.u.) & $RDEF$(fm)  \\[2.0ex] % inserts table
%heading
\hline\\[-2.0ex] % inserts single horizontal line
3/2$\leftrightarrow$5/2 & 57.99 & 365 & 3.75   \\ % inserting body of the table
5/2$\leftrightarrow$7/2 & 79.51 & 240 & 3.52    \\
3/2$\leftrightarrow$7/2 & 137.5 & 142 & 2.71   \\
7/2$\leftrightarrow$9/2 & 103.6 & 118 & 2.76   \\
5/2$\leftrightarrow$9/2 & 183.1 & 220 & 3.77   \\[0.7ex]
 % [1ex] adds vertical space
\hline %inserts single line
\end{tabular*}
\label{table:t2} % is used to refer this table in the text
\end{table}

\begin{figure}[t]
	\centering
    \includegraphics[clip, trim=0.0cm 0.0cm 0.0cm 0.0cm,width=0.48\textwidth]{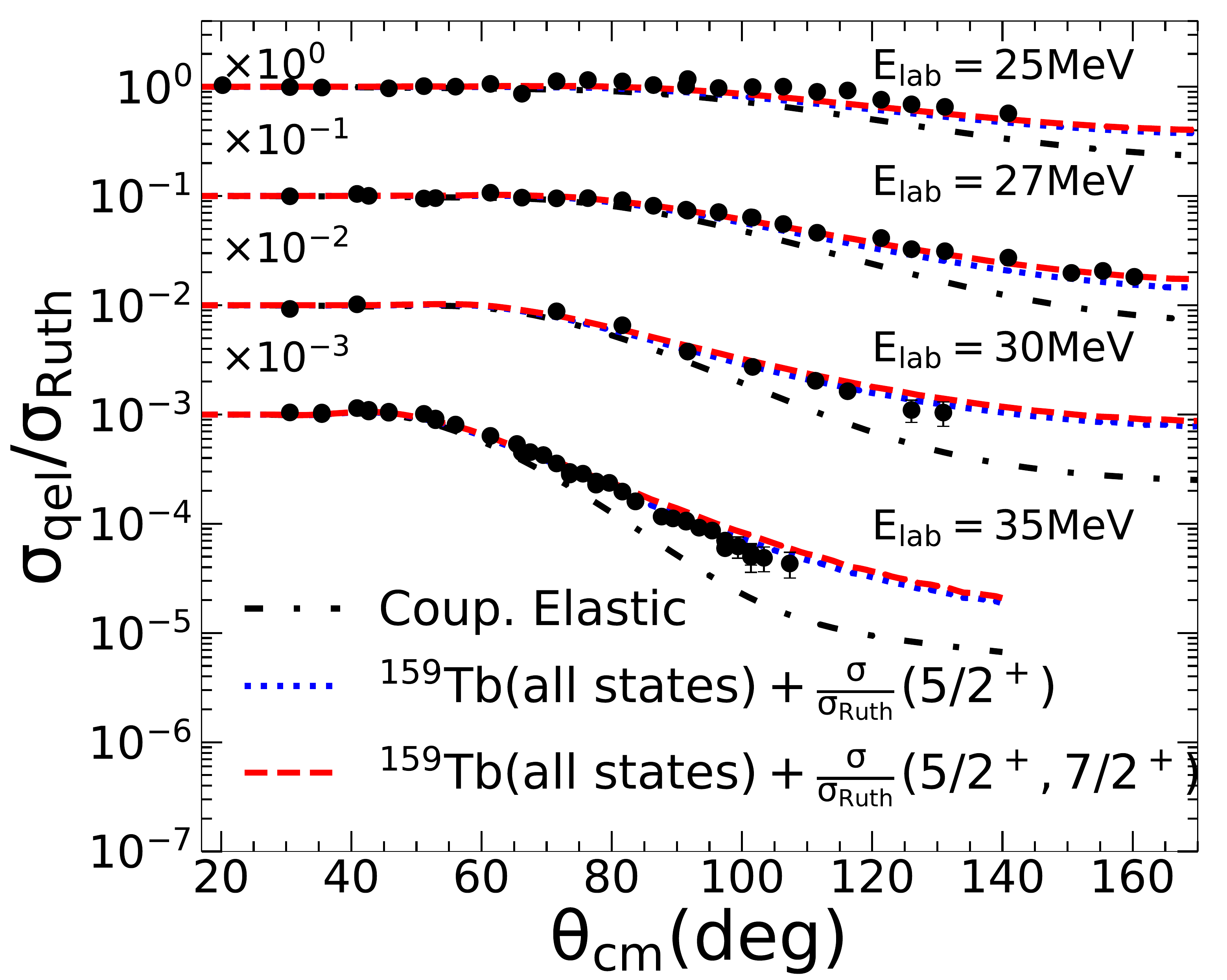}
  	\caption{Quasielastic scattering angular distribution for $^6$Li+$^{159}$Tb system at different energies. The experimental data has been compared with CC calculations. The addition of first excited state (5/2$^+$) to the elastic scattering cross-sections was required to reproduce the experimental quasielastic scattering cross-section. Incorporating the second excited state (7/2$^+$) did not change the theoretical cross-sections significantly. Higher excited states of $^{159}$Tb has even smaller contributions to the overall cross-sections, and thus were not included in the figure.} 
\end{figure}

For CC calculations, experimental reduced transition probabilities  \textit{B(E2)}  for different excited states of the target $^{159}$Tb  \cite{Reich12} has been used as input parameters. Collective excitation of target within rotational model has been considered. The code FRESCO takes \textit{B(E2)} value in the unit of $e^2fm^4$  which is related to the reduced transition strength in Weisskopf Unit (W.u) as:
\begin{align*}
B(E2)_{W.u.}= 0.05940A^{4/3} e^2fm^4 \tag{2}
\end{align*}
For the present calculation, g.s. and first two excited states of $^{159}$Tb have been included. The reduced nucelar deformation length for the transitions ($RDEF$) have also been calculated in the rotational model, and included as input parameters. Table III consists of the experimental B(E2) values and calculated $RDEF$ for $^{159}$Tb. 

\subsection{Comparison with quasielastic scattering angular distribution}
The effect of coupling on quasielastic scattering angular distribution has been shown  in Figure 6. The cross-sections  of the first two excited states at different angles obtained from CC calculations, have been added with the elastic scattering cross section angular distribution. The addition of first excited (5/2$^+$) state is required to reproduce the experimental quasielastic scattering cross section. The addition of the second excited state does not change the cross-section significantly. The third excited state is 241 keV apart, and the $\Delta$E-E detectors should be able to resolve any contribution occurring from that state (resoultion of $\Delta$E-E detectors are $\approx$ 150 keV). So, the cross-section of third excited state was not explicitely added. The addition of inelastic scattering cross-sections of the first two excited states was required to reproduce the experimental quasi-elastic scattering cross-section. 

\subsection{Comparison with inelastic scattering angular distribution}
The inelastic scattering from first excited state (58 keV) that was extracted from the quasielastic band (see Figure 2) has been compared independently with that calculated by CC calculation in Figure 7, at different energies. However, the extraction process becomes more intractable with decreasing energies for reasons described in section II, especially at E$_{lab}$=  25 MeV. At this energy, comparison between the CC calculation and extracted inelastic cross section was not attainable for a broad angular range. At the higher energies, such comparison yields an overall satisfactory result. The contribution of second excited state could not be compared in similar procedure as that part could not be extracted from experimental data. The contribution of second excited state can be observed in Figure 6.
 
\begin{figure}[t]
	\centering
    \includegraphics[clip, trim=1.0cm 0.0cm 0.0cm 0.0cm,width=0.49\textwidth]{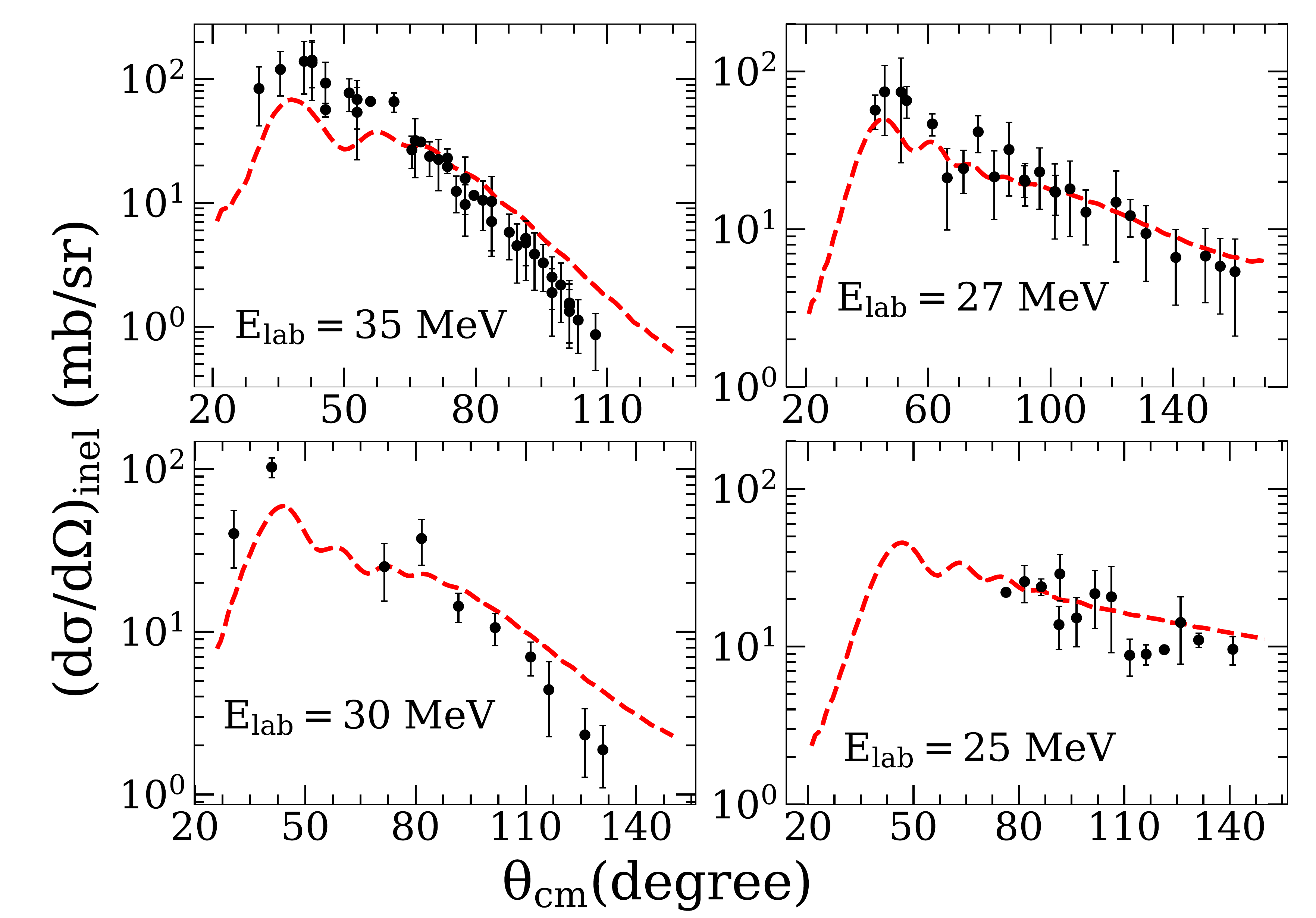}
  	\caption{Inelastic scattering angular distribution for $^6$Li+$^{159}$Tb (5/2$^+$; 58 keV) at different energies. The extracted inelastic cross sections are shown by dots, along with estimated errors. The dashed line represents the inelastic scattering cross section angular distribution obtained from CC calculation. } 
\end{figure}
\subsection{Finite range coupled channel Born approximation}
For transfer reactions of type $A + B (B'+X) \rightarrow (A+X) +B'$; (where $X$ is the particle transferred from $B$ to $A$), the theoretically calculated differential cross-section $\frac{d\sigma}{d\Omega}_{th}$ is related to the experimental cross-section as:
\begin{align*}
\frac{d\sigma}{d\Omega}_{exp}= (C^2 S)_{AX}(C^2 S)_{B'X} \frac{d\sigma}{d\Omega}_{th} \tag{3}
\end{align*}
where, $C^2$ is the isospin Clebsh-Gordan coefficients and $S$ is the spectroscopic factor of the cluster configurations. The product $C^2 S$ bears information regarding the probability of a nucleus to be found in a specific configuration. The framework of finite range coupled channel Born approximation (CCBA) \cite{Tamura74} has been chosen to describe the 1 neutron (\textit{1n}) pick up reaction by $^6$Li from well deformed $^{159}$Tb target nucleus. The rationale behind the choice of the model is the significant probability of two step processes, like one of the nuclei in the incident channel get inelastically excited followed by particle transfer from the state.

\begin{figure}[t]
	\centering
    \includegraphics[clip, trim=2.0cm 2.0cm 0.0cm 2.0cm,width=0.48\textwidth]{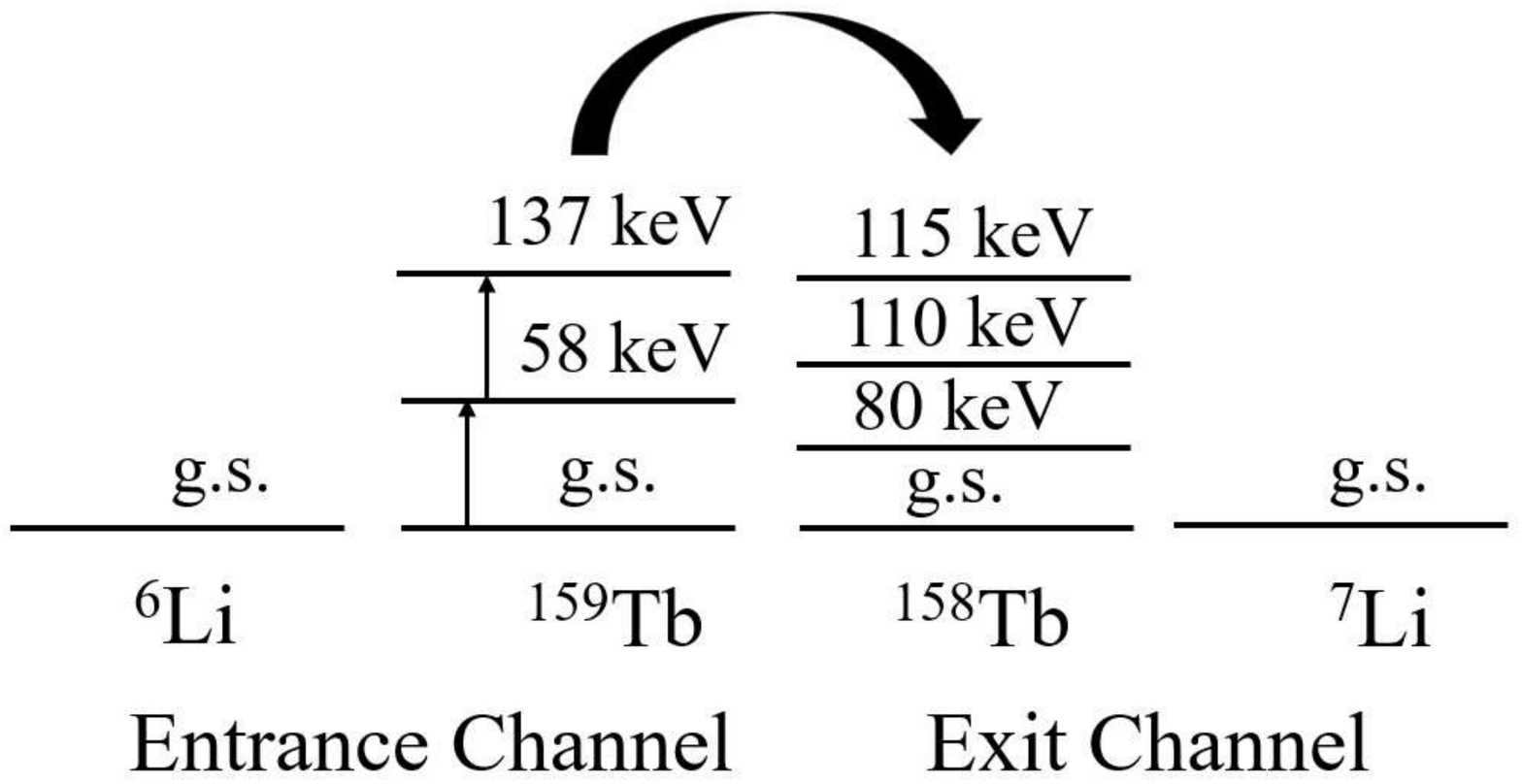}
  	\caption{Schematic diagram of the CCBA calculation performed for $^6$Li+$^{159}$Tb$^*$ $\rightarrow$ $^7$Li+$^{158}$Tb$^*$ reaction. See text for detailed description.} 
\end{figure}

\begin{figure}[b]
	\centering
    \includegraphics[clip, trim=0cm 0cm 0.0cm 0cm,width=0.49\textwidth]{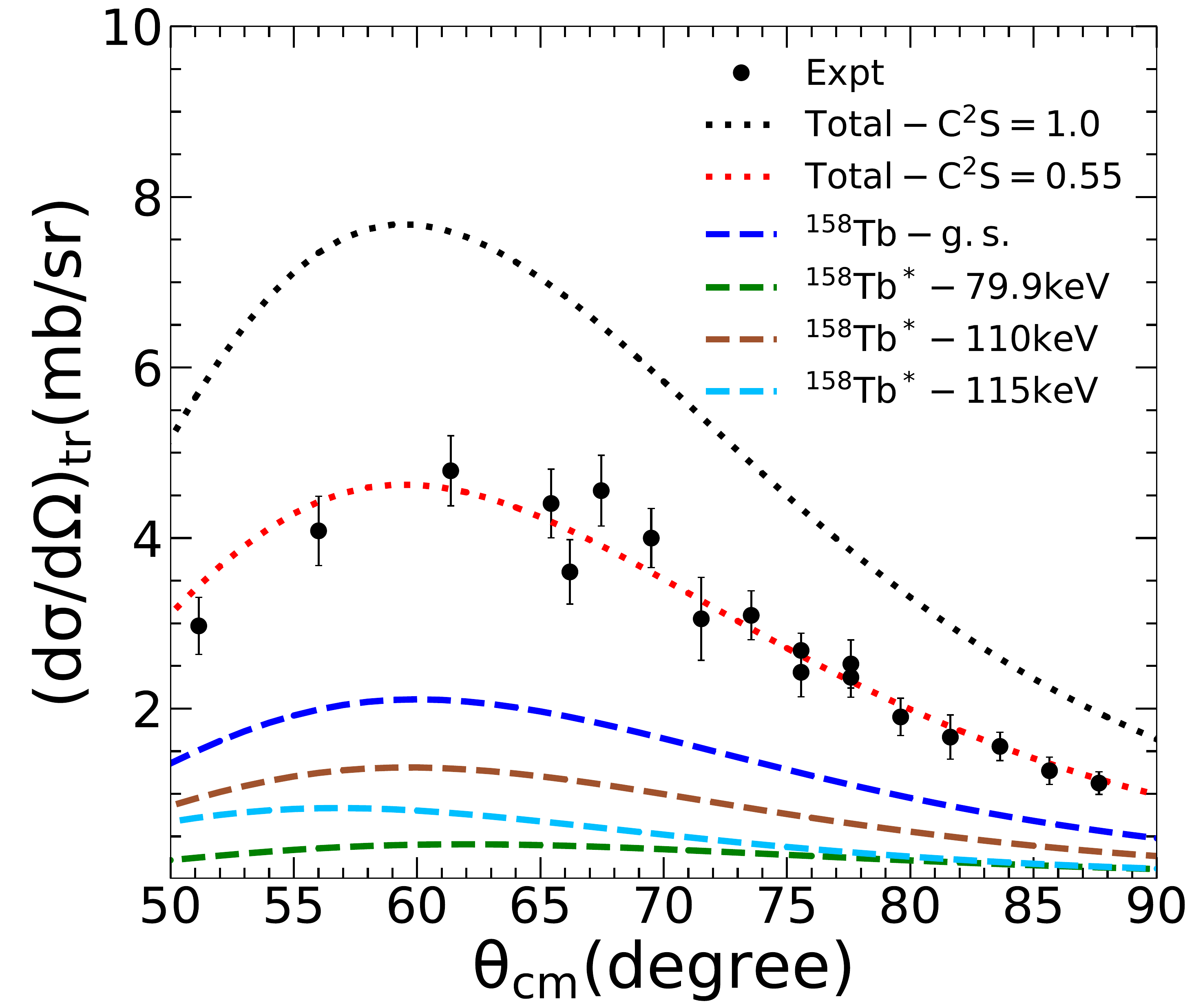}
  	\caption{Experimental cross-section angular distribution for $^{159}$Tb$^*$($^{6}$Li,$^{7}$Li)$^{158}$Tb$^*$ (black dots) at E$_{lab}$ = 35 MeV is fitted with individual theoretical cross-section occurring from different residual states of $^{158}$Tb. All of the individual theoretical cross-sections have been calculated with $C^2 S$=1.0 (dotted and dashed-dot lines). The best fit (black solid line) provides the estimated $C^2 S$ values of different spectroscopic configurations for $^{159}$Tb$\rightarrow$ $^{158}$Tb+n bound state.} 
\end{figure}

In the present work, CCBA calculation on 1$n$ pickup by $^6$Li from the $^{159}$Tb target nucleus has been performed. The experimental cross-sections were obtained from the $^7$Li band in the 2D spectrum shown in Figure 1. The Q-value of the reaction and the kinematical calculation suggest that the $^7$Li band is within the energy window expected from the \textit{1n} pickup by $^6$Li. The $^7$Li band was distinguishable for E$_{lab}$=27, 30 and 35 MeV. 
\begin{table}[t]
    \caption{$J^\pi$ and energy values  of the ground and excited states for $^{158}$Tb nucleus which have been used for calculation.}
    \begin{tabularx}{1.0\columnwidth}{XXX}
        \hline\hline\\ [-2.0ex] 
        $J^\pi$ & & E (keV)  \\[0.6ex]
        \hline
3$^-$ & & g.s.    \\
4$^-$ & & 79.9         \\
0$^-$ & & 110    \\
1$^-$ & & 115          \\

        \hline
    \end{tabularx}
\end{table}
\begin{figure}[t]
	\centering
    \includegraphics[clip, trim=0.0cm 0.0cm 0.0cm 0.0cm,width=0.48\textwidth]{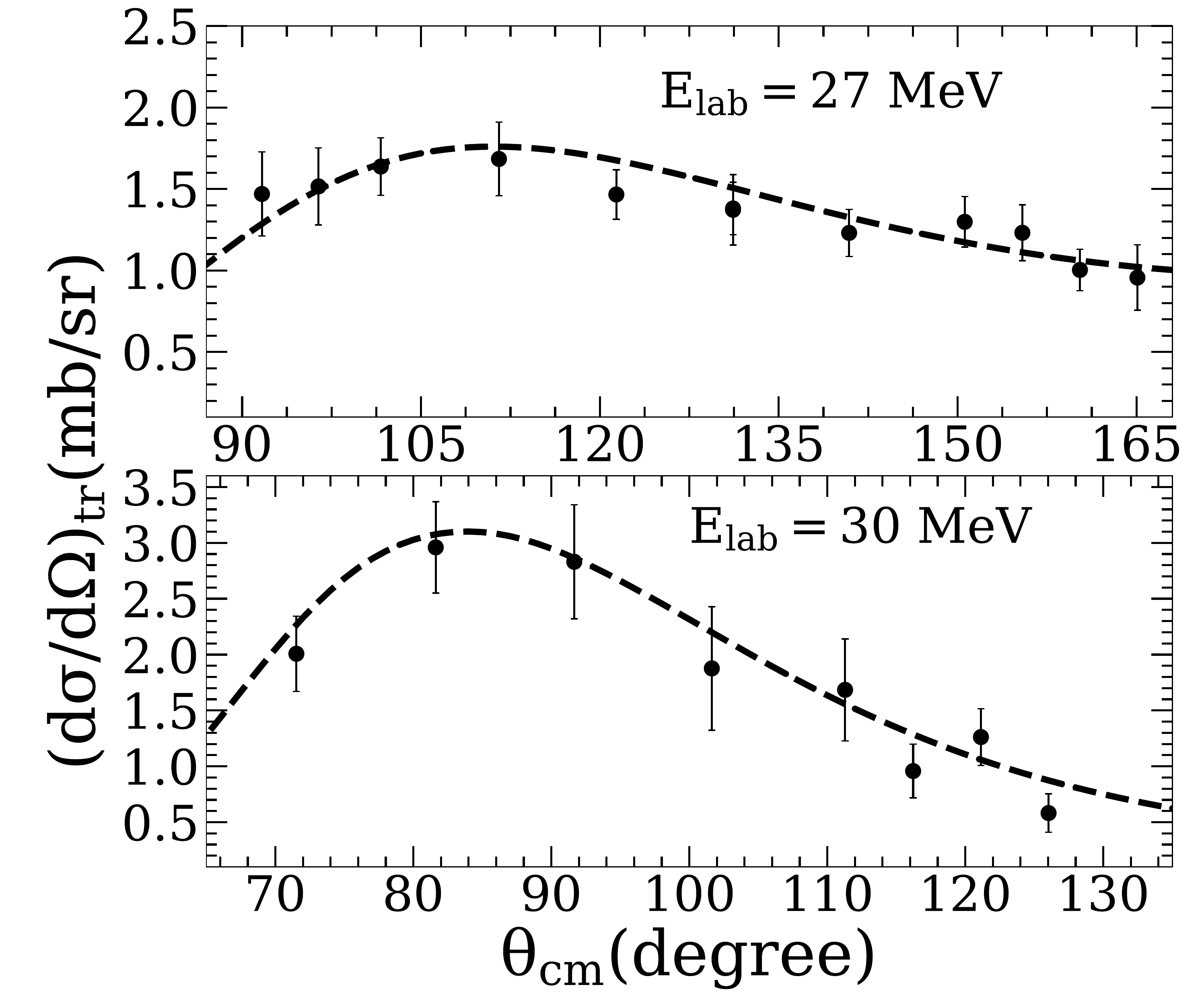}
  	\caption{Experimental cross section for 1n pickup reaction $^{159}$Tb$^*$($^{6}$Li,$^{7}$Li)$^{158}$Tb$^*$ (black dots) at E$_{lab}$= 27 and 30 MeV; compared with CCBA calculations (black dashed lines); using the  spectroscopic factors in Table V and Table VI. } 
\end{figure}
\begin{table}[t]
   \caption{Spectroscopic factors of the possible bound state configurations for $^7$Li and $^{159}$Tb [$^{158}$Tb(3$^-$,g.s.)+n]. Each partition in the table is individually  employed for CCBA calculation. $C^2 S_\mathrm{^{158}Tb+n}$ was initially fixed at 1.0. The best fit provided the estimated value of $C^2 S_\mathrm{^{158}Tb+n}$ (in column 8). }
    \begin{tabularx}{1.0\columnwidth}{XXXX}
        \hline\hline\\ 
        Nucleus & E$_x$(keV) & State ($nlj$) &$C^2 S$(this work) \\[1.0ex]
        \hline\\[-1.3ex]
         $^{158}$Tb & g.s.& $^3p_{3/2}$ & 0.55$\pm$0.05\\[1.0ex]
         & & $^2f_{7/2}$ & 0.55$\pm$0.05\\[1.0ex]
         & & $^2f_{5/2}$ & 0.55$\pm$0.05\\[1.0ex]
         & 79.9 & $^2f_{5/2}$ & 0.55$\pm$0.05\\[1.0ex]
         & 110.0& $^3p_{3/2}$ & 0.55$\pm$0.05\\[1.0ex]
         & & $^2f_{5/2}$ & 0.55$\pm$0.05\\[1.0ex]
         & & $^2f_{7/2}$ & 0.55$\pm$0.05\\[1.0ex]
         & 115.0& $^3p_{3/2}$ & 0.55$\pm$0.05\\[1.0ex]
         \hline

    \end{tabularx}
\end{table}

A schematic diagram of the coupling scheme for the CCBA calculation is presented in Figure 8. The first and second inelastic excitations of $^{159}$Tb nucleus along with its g.s. are first coupled with $^{6}$Li g.s. as described previously. The third excited state at 241.5 keV has small contributon on transfer cross-sections. Therefore, it was not included during transfer calculations. Following the reaction  $^{159}$Tb$^*$($^{6}$Li,$^{7}$Li)$^{158}$Tb$^*$, the residual $^{158}$Tb nucleus can end up either in its g.s. or excited states (e.s.) \cite{Nica17}. The $J^\pi$ values and energies of those states are listed in Table IV. For a comprehensive analysis, cross-sections of the g.s.and the e.s. of $^{158}$Tb need to be calculated. In addition, five sets of potential are required for calculation, listed below:

(i) Entrance channel $^6$Li+$^{159}$Tb potential : obtained from calculations in previous section.

(ii) Exit channel $^7$Li+$^{158}$Tb potential: obtained from Ref. \cite{Patel15}. These potentials are originally for $^7$Li+$^{159}$Tb system, but due to unavailability of $^7$Li+$^{158}$Tb potential, the parameters from Ref. \cite{Patel15} have been used.

(iii) Neutron bound state of target ($^{158}$Tb+n): potential parameters of Ref. \cite{Becchetti69} used.

(iv) Neutron bound state of projectile ($^{6}$Li+n): pot. parameters obtained from Ref. \cite{Li69}.

(v) core-core potential ($^6$Li+$^{158}$Tb): The parameters obtained in previous sections used. The actual "$\textit{core-core\ interaction\ potential}$" might differ from the used value but any such potential parameters are unavailable in literature. The parameters have been altered to extremes ($V_0$ and $W_0$ changed from 10-200 MeV) to check its sensitivity on calculated cross sections. The cross sections did not alter to a significant extent, and nature of the cross-section angular distributions occurring from different partitions remained same. 

The spectroscopic factor ($C^2 S$) for $^6$Li+n configuration of $^7$Li is found in literature \cite{Rudchik14}. But, in view of the unavailability of $C^2 S$ values for $^{158}$Tb+n configuration, the same values have been obtained at first, from the experimental cross-section at E$_{lab}$=35 MeV.  Theoretical calculations have been performed at that energy for different spectroscopic configuration of $^{158}$Tb+n bound states; exclusively, each with initial $C^2 S$=1.0. The possible configuration for  $^{158}$Tb+n bound state is presented in Table V with their respective $C^2 S$ values. Performing CCBA for $^{159}$Tb$^*$($^{6}$Li,$^{7}$Li)$^{158}$Tb(g.s.) was not sufficient to reproduce the experimental data. Thus the $^{159}$Tb$^*$($^{6}$Li,$^{7}$Li)$^{158}$Tb$^*$(79.9, 110.0, 115.0 keV) coupling schemes were also included in the calculation. Figure 9 represents the experimental and theoretical differentail scattering cross-section for 1n pickup ($\mathrm{(d\sigma/d\Omega)_{tr}}$) at E$_{lab}$ = 35 MeV. Experimentally, it was not possible to distinguish the different states of $^{158}$Tb$^*$. Hence, a single $C^2 S$ = 0.55$\pm$0.05 was used for each state to match the experimentally obtained 1$n$ pickup cross-section.

 CCBA calculations at E$_{lab}$= 27 and 30 MeV was performed with the obtained $C^2 S_\mathrm{^{158}Tb+n}$ values and compared with the experimental data in Figure 10. Good results have been obtained in both energies, which can be considered as the credibility of the obtained values. Below 27 MeV, the $^7$Li band could not be separated from $^6$Li band, experimentally, prohibiting further study of the below barrier 1n pickup reaction.    

\section{Summary}
Quasielastic scattering angular distribution have been measured at energies above and below the Colulomb barrier for $^6$Li+$^{159}$Tb system. The pure elastic scattering cross-sections have been extracted from the quasielastic cross-sections. The optical model potential parameters have been obtained from the extracted elastic scattering cross-section angular distribution, at each energy. Coupled channel calculation has been performed, including the coupling with first three excited states of $^{159}$Tb target, and no projectile excitation. The extracted optical model potential parameters are used for calculation. Reasonable agreement has been achieved between experimental and thoeretical quasielastic and inelastic cross sections.

The cross-sections for \textit{1n} pickup reaction for this system has been studied. The $^{159}$Tb $\rightarrow$ $^{158}$Tb+n spectroscopic factors have been estimated from E$_{lab}$ = 35 MeV angular distribution data, with the aid of CCBA calculation. The calculated energy variation of \textit{1n} pickup cross-sections agree well with that of the experimental cross-sections. To match the theoretical and experimental cross-sections, a C$^2$S = 0.55 $\pm$0.5 was required. Overall, the quality of fits to the reaction channels within the coupled channel framework highlights the correctness of the extraction of the elastic cross section and the resultant optical model parameters for $^6$Li+$^{159}$Tb system. Experimental measurement spanning more energy range can return a comprehensive result regarding values of different parameters, estimated in this work.

\begin{acknowledgments}
The authors sincerely thank the staff of BARC-TIFR Pelletron facility for an uninterrupted supply of beam. One of the Authors, S.B. would like to thank Prof. S. Kailas of CBS, Mumbai and Dr. Rajkumar Santra of VECC, Kolkata; for discussion on theoretical frameworks that have been employed in this work. 
\end{acknowledgments}
 
%\vspace{5mm}

\end{document}